\documentclass{vldb}

\usepackage{graphicx}
\usepackage{balance}
\usepackage{amsmath}
\usepackage{amssymb}

\usepackage{framed}

\newcommand{\rb}{\ensuremath{\mathbf}}

\begin{document}

\title{80 New Packages to Mine Database Query Logs\titlenote{Number of 
        R packages for machine learning recommended by the official website (December
        2015)\\\texttt{cran.r-project.org/web/views/MachineLearning.html}}}

\numberofauthors{2}
\author{
\alignauthor
Thibault Sellam\\
\affaddr{CWI, the Netherlands}\\
\email{thibault.sellam@cwi.nl}
\alignauthor
Martin Kersten\\
\affaddr{CWI, the Netherlands}\\
\email{martin.kersten@cwi.nl}
}

\maketitle
\thispagestyle{empty}

\begin{abstract}
The query log of a DBMS is a powerful resource. It enables many practical
applications, including query optimization and user experience enhancement.
And yet, mining SQL queries is a difficult task. The fundamental problem is that
queries are symbolic objects, not vectors of numbers.  Therefore, many popular
statistical concepts, such as means, regression, or decision trees do not
apply.  Most authors limit themselves to ad hoc algorithms or approaches based
on neighborhoods, such as $k$ Nearest Neighbors.  Our project is to challenge
this limitation. We introduce methods to manipulate SQL queries as if they were
vectors, thereby unlocking the whole statistical toolbox. We present three
families of methods: feature maps, kernel methods, and Bayesian models.  The
first technique directly encodes queries into vectors.  The second one transforms
the queries implicitly. The last one exploits probabilistic graphical
models as an alternative to vector spaces. We present the benefits and
drawbacks of each solution, highlight how they relate to each other, and make
the case for future investigation.
\end{abstract}

\section{Introduction}
\label{sec:intro}

The query log of a SQL database gives us precious hints about what its users are
interested in. From this dataset, we can infer query
auto-completions~\cite{akbarnejad2010sql, khoussainova2010snipsuggest,
sarawagi1998discovery}. We can simulate realistic queries, for testing
purposes~\cite{tran2015oracle}. We can even reduce the latency of the queries, thanks
to speculative execution~\cite{sapia2000promise}. Furthermore, the log
describes the database itself: it describes which queries succeeded or failed,
how long they took, and how many tuples they returned. Combined with predictive
algorithms, this information could help us emit warnings, chose efficient query
plans and build more robust engines.

Yet, mining query logs is subject to a fundamental problem: SQL queries do not
live in a vector space. In their na\-tural form, queries are structured, symbolic
objects - not vectors of real numbers.  Hence, the vast majority of statistical
concepts are undefined. Elementary methods such as means, correlations or
regression do not apply. The same problem arises with advanced methods such as
neural networks or SVMs. Consequently, most authors resort to
\emph{appli\-ca\-tion-specific frameworks}~\cite{agrawal2006context, ghosh2002plan,
giacometti2009query, yang2009recommending, yao2005finding, zhang2011data}: they
devise some encoding specifically for the problem at hand, and feed
it to a custom algorithm.  This approach is neither practical nor efficient,
because each use case requires a complete new representation system and a new
algorithm.

A few authors have developed more general, application-independent solutions:
\emph{neighborhood-based algorithms}~\cite{akbarnejad2010sql,
aligon2014similarity, chatzopoulou2009query, Nguyen2015Ident}. These algorithms
are popular because they require no encoding. Instead, they rely on a
\emph{pairwise dissimilarity function}, which quantifies the similarity or
difference between two queries. Once the authors have defined such a function,
they apply it to all the pairs of queries in the log. They obtain a
neighborhood graph, in which they detect discrete patterns.  But these methods
are limited: we observed that few papers, if any, venture beyond the strict
realm of clustering and $k$ Nearest Neighbors (kNN). One explanation is that
statistical textbooks and software provide little support for other tasks.  To
illustrate, the official R Website does not even mention NN-regression on its
machine \mbox{learning} page (cf. footnote). Besides, these approaches suffer from
quali\-tative drawbacks.  They cannot interpolate between training examples,
e.g., to compute centroids.  They have little to no notion of prediction
confidence.  Finally, they are very sensitive to small training sets, local
sparsity, and class imbalance. Several empirical studies reveal cases where
they are under-optimal~\cite{desrosiers2011comprehensive,
koren2008factorization}.

Our ambition is  to unlock the rest of the statistical toolbox.  We want to
perform kNN and clustering, but also density estimation, sampling, regression,
classification, dimen\-sionality reduction, reinforcement learning and
visualization, directly over SQL queries. To do so, we develop me\-thods to
encode the query log in such a way that it becomes subject to these tasks.
We envision software ``converters'', to process query logs in R, Weka or Matlab
as if they were classic tables of numbers.  Thus, database designers will
be\-nefit from the rich libraries offered by these platforms. They will be able
to focus on insights and functionalities rather than implementation.
\pagebreak

In this paper, we describe promising methods to represent query logs in an
application-independent fashion. We present three families of encodings:
\begin{itemize}
    \item \emph{Feature maps} directly transform queries into vectors.
    \item \emph{Kernel methods} manipulate queries as if they were vectors, but
        without actually transforming them.
    \item \emph{Bayesian methods} rely on probabilistic graphical mo\-dels rather than
       vector spaces.
\end{itemize}
We highlight the advantages and drawbacks of each solution, and present
mathe\-matical transformations to switch from one representation to the other.
For all three families, we make the case for longer term investigations.

The rest of the paper is organized as follows. In Section~\ref{sec:motivation},
we motivate our work and we present our requirements.  In
Sections~\ref{sec:explicit}, \ref{sec:kernel} and \ref{sec:generative}, we
introduce our solutions. We highlight their relationships in
Section~\ref{sec:bridge}. We discuss related work in
Section~\ref{sec:RelatedWork}, and conclude in Section~\ref{sec:conclusion}.

\section{Overview}
\label{sec:motivation}

We established that queries do not live in a vector space. But what if we could
devise a function $\Phi$ to transform SQL statements into vectors? In this
section, we present the immense range of practical applications which would
follow.  We then discuss how realistic this vision is.

\subsection{Visions for Query Log Mining}
\label{sec:visions}

\begin{figure}[!t]
  \centering
    \includegraphics[width=\columnwidth]{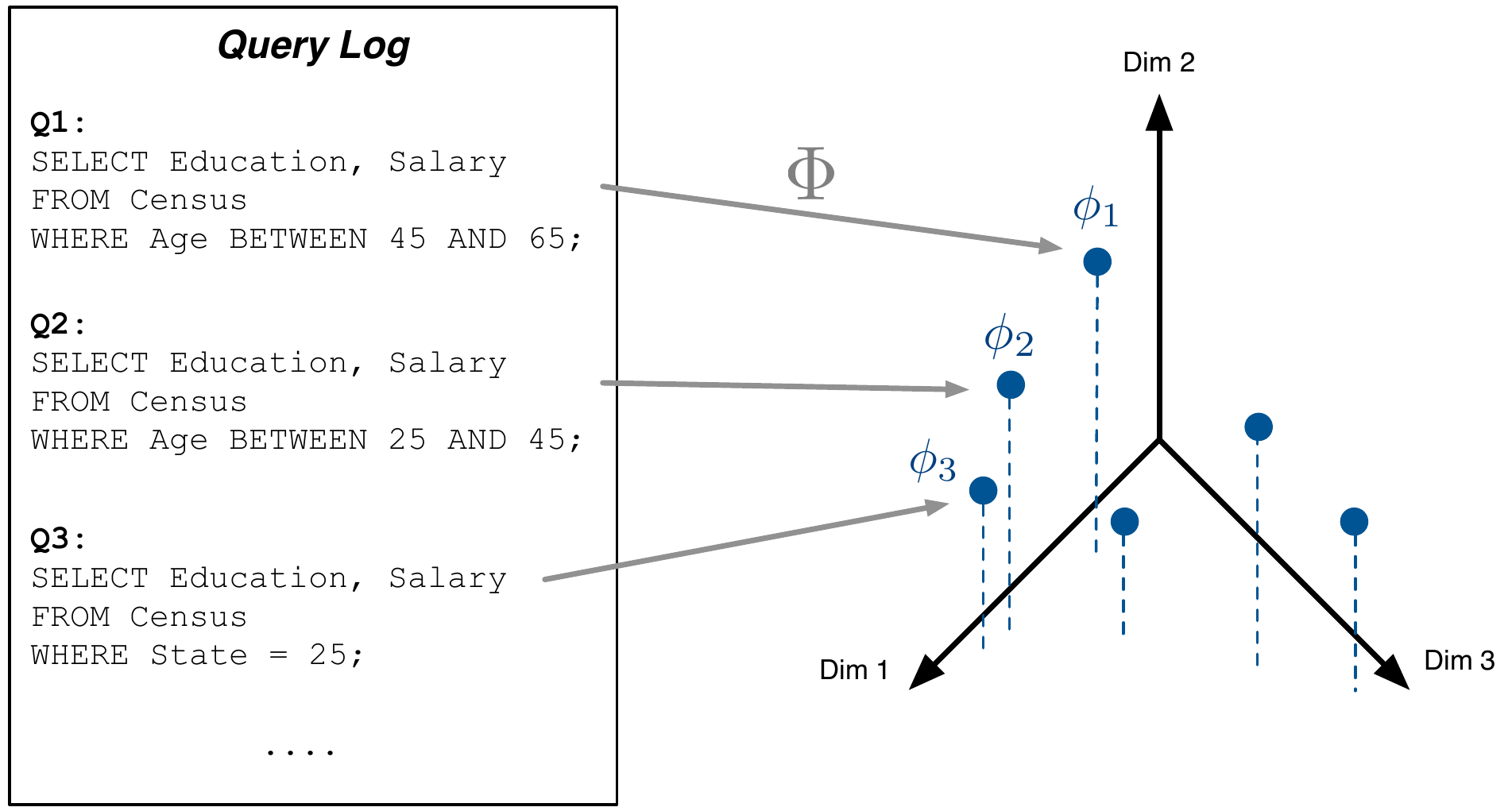}
  \caption{Example of feature map $\Phi$.}
  \label{pic:featuremap}
\end{figure}

\textbf{From Queries to Vectors.}
Suppose that we could access a function $\Phi$, to map any SQL query $Q \in
\mathbb{Q}_{SQL}$ to a vector $\phi \in \mathbb{R}^D$. We illustrate it in
Figure~\ref{pic:featuremap}.  To be consistent with the machine learning
literature, we name it \emph{feature map}~\cite{bishop2006pattern}, and we suppose
that it is one-to-one. How could this function be useful?

\begin{figure}[!t]
  \centering
    \includegraphics[width=.9\columnwidth]{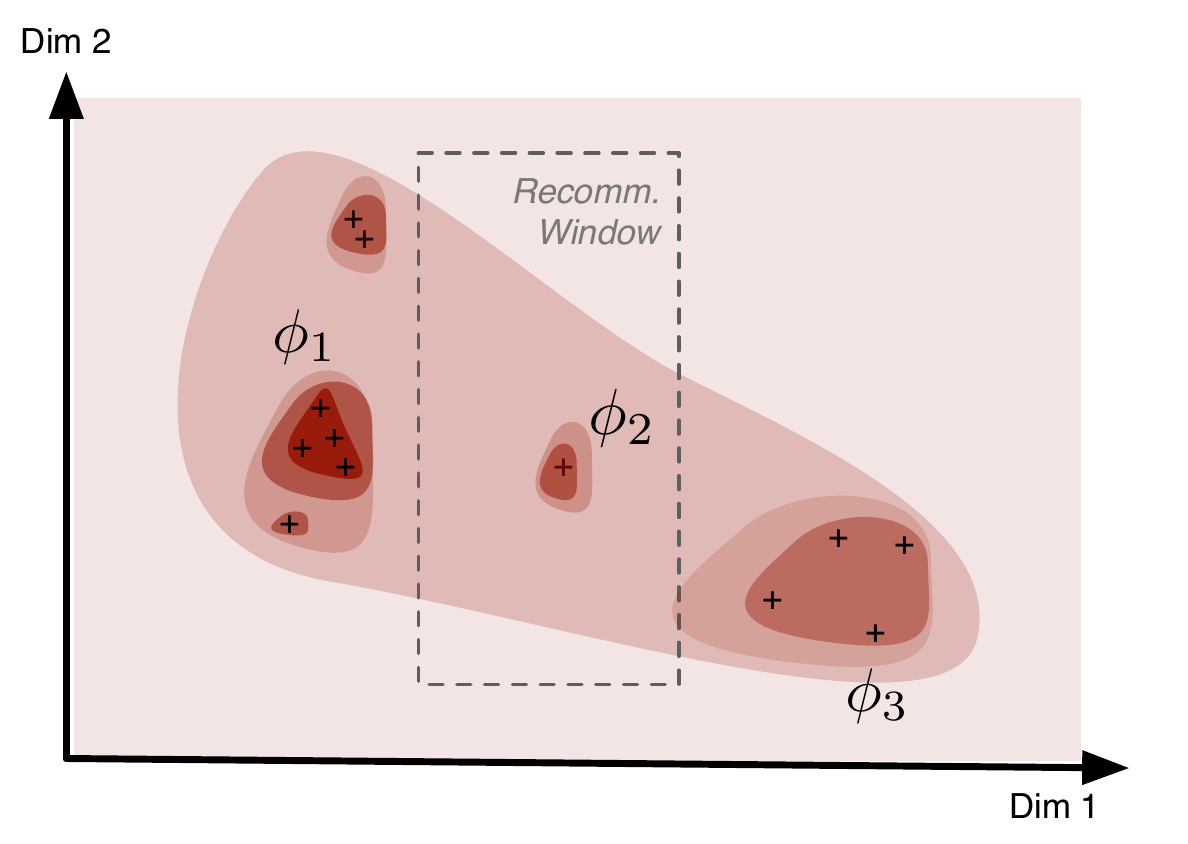}
    \caption{Heatmap of the log's density. The dashed rectangle
    represents the recommendation window, defined by the user's input.}
  \label{pic:hotzones}
\end{figure}
First, we could perform \emph{density estimation}: for each query $Q$, we could
estimate the probability function $p(\Phi(Q))$, as illustrated in
Figure~\ref{pic:hotzones}. The density function is a powerful tool, because it
lets us perform many classic tasks from the log mining literature.  For
instance, we could detect ``hot zones'' in the query log (i.e. clusters). We
could also re\-commend queries: when users start typing SQL statements, they
implicitly define a window of interest, as shown in Figure~\ref{pic:hotzones}.
To help them, we could highlight the most popular queries in this window.

More importantly, a function $\Phi$ would allow us to perform
\emph{regression} and \emph{classification}. In regression, we infer
quantities from SQL statements, based on past observations.  Thanks to this
method, we could estimate the runtime of a query, the cardinality of its
output, or or the number of machines involved in a cluster. In
classification, we predict a discrete variable. Thus, we
could detect which user is currently querying the database, and pre-fetch some
data accordingly. We could also emit warnings, if the user's query is
dangerously close to one that failed previously. Finally, we could
machine-learn tasks which were previously coded by hand: among others, we could
train a neural network to associate SQL queries with visualizations.

To conclude, the combination of the function $\Phi$ and statistical algorithms
would lead to dozens of applications. A few of them have been proposed in the
literature before (those related to density estimation), others are new. In any
case, they would all run on top of a unified, complete formalism.

\textbf{From Vectors to Queries.} We now go one step further: what if we had
access to an \emph{inverse feature map}  $\Phi^{-1}$ to reconstitute queries
from vectors?

The function $\Phi^{-1}$ would have a dramatic effect: it would let us create
new queries from scratch. Observe the density function pictured in
Figure~\ref{pic:hotzones}. By \emph{sampling} from this distribution, we could
produce queries that have never been written before, but which are likely to
occur. Thus, we could generate artificial, but realistic workloads. This
technique could be useful for testing and exploration.  Combined with adaptive
indexing mechanisms such as database cracking~\cite{halim2012stochastic}, it
could also help us build more efficient indices.

\begin{figure}[!t]
  \centering
    \includegraphics[width=0.95\columnwidth]{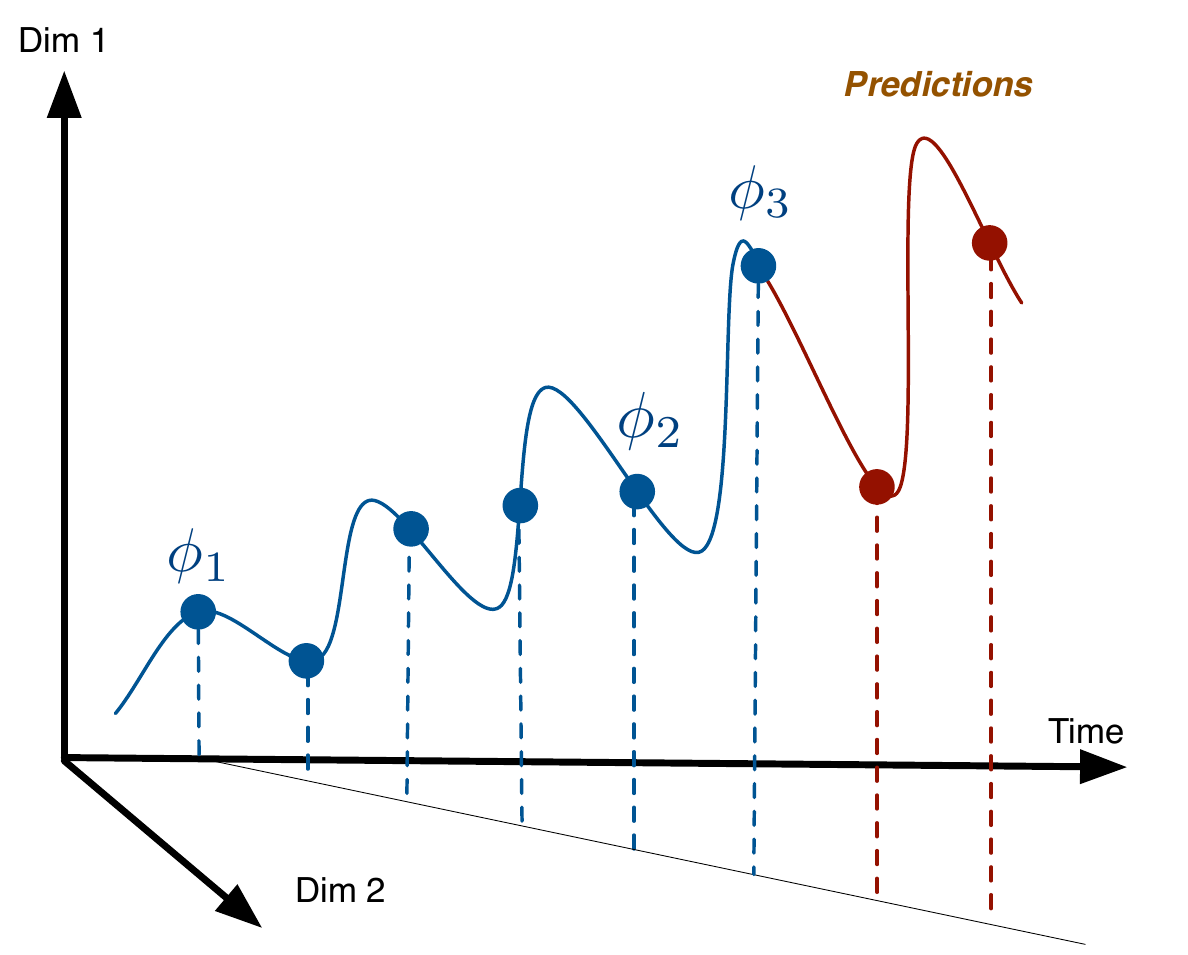}
    \caption{Example of querying pattern based on time.}
  \label{pic:timeseries}
\end{figure}

Another application of this idea is \emph{query regression}: we could
extrapolate SQL queries from other SQL queries. Consequently, we could detect
usage patterns, and exploit those to predict which query will come next, using
time series models.  Figure~\ref{pic:timeseries} provides an example.  This
scenario is fictive, and we suspect that real workloads are more chaotic in
practice.  But we do not need to predict precise queries. Predicting general
areas of interest would already be helpful, and probabilistic methods excel at
that.

Finally, more applications could come from \emph{active learning}. In
particular, we envision adaptive DBMS benchmarks. Such systems would pose
queries, observe how the database reacts and adapt their behavior accordingly.
Thus, they would automatically identify performance bottlenecks, and report them
to DBMS designers.

\subsection{How Far Are We?}
\label{sec:discussion}

In fact, constructing a function to map queries to vectors is not a difficult
task. For example, we could count $n$-grams, as in information retrieval. The
whole challenge is to build an application-independent transformation. Such a
transformation should be \emph{lossless}, that is, non destructive.  The vector
representation of a query should convey all the information contained in its
SQL form. It should contain lexical and grammatical information: which keywords
are used, and what are their roles. But it should also convey the \emph{set
relationships} between the queries. By nature, queries represent sets of
tuples, which can be disjoint, overlapping, or nested.  With continuous
variables, they can even be ordered. These properties should be preserved
in the encoding. The actual feature selection, which depends on the use case,
should be left to the user.

Unfortunately, we suspect that if such a mapping $\Phi$ exists, then the vector
space it yields will have a huge, unpractical dimensionality. We  discuss this
point further in Section~\ref{sec:explicit}. In the rest of this paper, we
present several restricted versions of the function $\Phi$.  Two of these
methods are lossless: dummy coding and Bayesian modeling.  However, their scope
is limited: we have not yet found any practical way to process all the possible
queries from SQL. The remaining approaches are more flexible, but they are
lossy. The users must specify the properties of interest in advance. For
instance, they may focus on the syntactical structure of the queries, or their
extent. The encoding will reflect these attributes, and destroy the remaining
information.  Consequently, two distinct queries can have the same encoding,
and the inverse mapping $\Phi^{-1}$ is undefined.

\section{Feature Maps}
\label{sec:explicit}
We now present two methods to build feature maps, \emph{dummy coding} and
\emph{dissimilarity-based feature maps} (DBFMs).

\subsection{Dummy Coding}
\label{sec:dummy}

\begin{figure}[!t]
  \centering
    \includegraphics[width=\columnwidth]{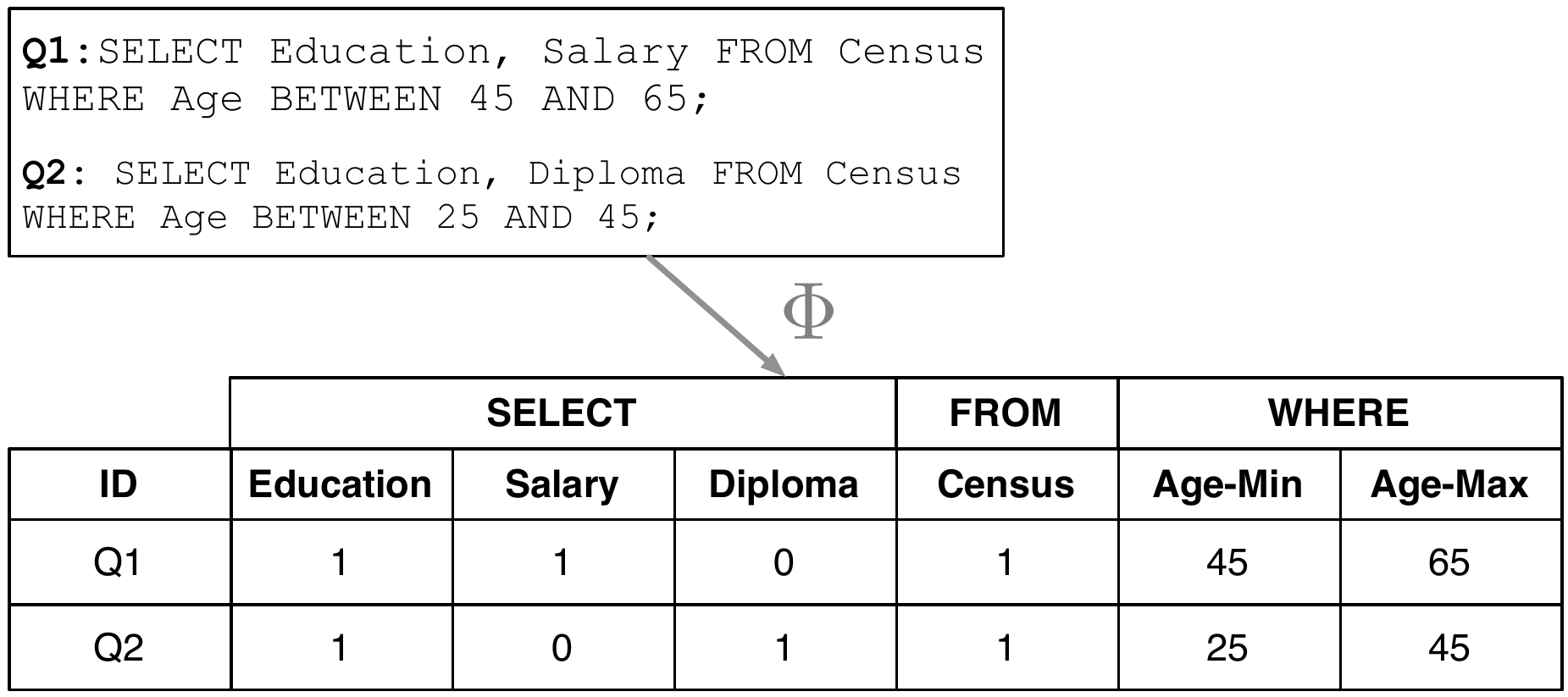}
    \caption{Example of dummy coding.}
  \label{pic:dummy}
\end{figure}

\textbf{Method.} The idea behind dummy coding is to represent queries with
vectors of binary variables, where each component represents a degree of
freedom offered by SQL. For example, a variable could signal the presence or
absence of a certain table in the \texttt{WHERE} clause, or an aggregation in
the \texttt{SELECT} section. Additionally, we include continuous columns to
deal with numeric selection predicates.  Figure~\ref{pic:dummy} illustrates
this method with a fictive example.

In fact, dummy coding has a fundamental flaw: to support all of SQL, it
requires vectors of infinite length. In consequence, we must limit its scope.
One option is to represent only the queries in the log, as we did in
Figure~\ref{pic:dummy}.  An other is to specify a subset of SQL a priori. For
example, we can restrict the encoding to Select-Project-Join queries with a
limited number of components.  Additionally, we can compress the resulting
vectors with dimensionality reduction methods, such as factor analysis or
autoencoders~\cite{bishop2006pattern}.

\textbf{Discussion.} Dummy coding is the naive approach. It is straightforward
and lossless. It produces flat tables, which effectively make it possible to
mine query logs with mainstream statistical tools. But we foresee that it will
return huge, sparse vector spaces with complex queries. The subsequent vectors
will be costly to store, to process, and statistical methods will be prey to
overfitting (as per the curse of dimensionality~\cite{bishop2006pattern}).
Dimensionality reduction algorithms can help, but they are lossy,  expensive,
and they require careful tuning.  Besides, binary variables are not real
numbers, thus not all statistical methods can cope with them (for example,
k-means is excluded). For all these reasons, we need alternative encoding
schemes.

\subsection{Dissimilarity-Based Mapping}
\label{sec:dbfm}
We now present dissimilari\-ty-based feature maps (DBFMs), which
generalize of existing work on query log mining.

\textbf{Method.}  To build a DBFM, we operate in three steps.  First, we chose
one or several pairwise dissimilarity measures from the literature. Second, we
embed them into an enco\-ding function. Thanks to this function, we can represent
the log with a large matrix. In the last step, we compress it.

Defining the dissimilarity between two queries is subject to all the problems
presented in Section~\ref{sec:discussion}. Currently, we know no perfect
measure of dissimilarity.  However, several authors have already proposed
specific functions, in the context of neighborhood-based approaches.
Chatzopoulou et al.~\cite{chatzopoulou2009query} have reported a measure based
on query results: two queries are similar if they involve the same tuples.
Akbarnejad et al.~\cite{akbarnejad2010sql} have used fragments of text.  More
recently, Nguyen et al.~\cite{Nguyen2015Ident} have developed a method to
exploit the results of queries without running them. In a recent paper, Aligon
et al. review 14 of these functions~\cite{aligon2014similarity}.
Collectively, those cover a wide range of use cases. Our idea is to embed them
in an encoding $\Phi$.

For a given dissimilarity measure, the square matrix
$\rb{D}$ represents the \emph{dissimilarity matrix} of the query log. This
matrix contains the pairwise dissimilarities between all the couples $(Q_i,
Q_j)$ in the log, as follows:
\begin{equation}
\rb{D} \equiv \begin{bmatrix}
    d(Q_1, Q_1) & d(Q_1, Q_2) & \ldots & d(Q_1, Q_N)\\
    d(Q_2, Q_1) & \ddots & &\vdots \\
    \vdots & & \ddots&\vdots \\
    d(Q_N, Q_1)&  \ldots & \ldots &d(Q_N, Q_N)\\
\end{bmatrix}
\end{equation}
It turns out that we can derive a trivial feature map from this representation:
we map each query $Q_i$ to the vector $\phi_i = [d(Q_i, Q_1), \ldots, d(Q_i,
Q_N)]^\top$. In other words, we  associate each query to its corresponding line
in $\rb{D}$. Hence, DBFMs represent queries by their difference with regards to
the other queries in the log. The resulting space is called \emph{dissimilarity
space}, and its theoretical properties were described by Pekalska and
Duin~\cite{duin2012dissimilarity}. Observe that this method lets us combine
several dissimilarity measures: we simply concatenate the resulting
dissimilarity matrices. To deal with the dimensions of the result, we apply
dimensionality reduction.  Specifically, we can use PCA, or we can cluster the
columns and pick a few representative dimensions.

\textbf{Discussion.} The advantage of the DBFM method is its flexibility.  In
comparison with dummy coding, DBFMs can deal with complex queries.  Also, they
generate continuous variables, which involves a broader class of algorithms.
However, these functions are lossy: the user must specify the properties of
interest. Also, the compression step is costly and it requires tuning, as
discussed in Section~\ref{sec:dummy}. Finally, DBFMs are by definition
sensitive to the  queries in the log.  If those are similar to each other, then
the columns of the dissimilarity matrix $\rb{D}$ will be highly correlated.
Therefore this matrix will contain little information.  The physical
dimensionality of the dissimilarity space will be high, but its intrinsic
dimensionality will be low. In conclusion, DBFMs appear as viable substitutes
for dummy coding in cases where the log is small and the queries diverse. But
we need more general methods for larger and sparser data sets.

\textbf{Multidimensional Scaling.} An alternative approach is Multidimensional
Scaling~\cite{borg2005modern}. This method takes the dissimilarity matrix
$\rb{D}$ as input, and generates a vector space in which the pairwise distances
between the objects are preserved. Multidimensional scaling is relevant, but it
suffers from the exact same problems as DBFMs: it is costly, it requires tuning
and it depends crucially on the queries in the log.

\section{Kernel Functions}
\label{sec:kernel}
In the previous section, we presented two general classes of feature maps.  We
now discuss \emph{implicit} alternatives: kernel approaches.

\subsection{Introducing Kernel Functions}
\label{sec:introkernel}

The aim of this section is to communicate the intuition behind kernels.  We
refer the reader to Bishop~\cite{bishop2006pattern} for a more rigorous
introduction.

In this paper, we mention a number of statistical methods applicable to
vectors, such as regression, classification and clustering. In fact, we do not
need all of algebra to perform them.  We need only one fundamental
operation: \emph{the dot-product}. If we can compute the dot-product $\phi_i
\cdot \phi_j$ between two vectors $\phi_i$ and $\phi_j$, then we can run linear
regression, Support Vector Machines, K-means, PCA and many others. The process
of rewriting a statistical method in terms of dot-products is known as
\emph{kernelization}~\cite{bishop2006pattern}.

At this point, computing the dot-product $\phi_i \cdot \phi_j$ is problematic
because we need to compute the vectors $\phi_i = \Phi(Q_i)$ and $\phi_i =
\Phi(Q_j)$. To do so, we need the mapping function~$\Phi$. Kernel functions
let us bypass this operation. A kernel function $K(Q_i, Q_j)$ is analog to a
dissimilarity measure: it has a low value if $Q_i$ and $Q_j$ are similar, and
it has a high value otherwise. But kernels have a convenient mathematical
property: for every such function $K$, there exists a feature map $\Phi$ such
that:
\begin{equation}
    K(Q_i, Q_j) = \Phi(Q_i) \cdot \Phi(Q_j)
\end{equation}
In plain words, computing the similarity between two queries according to $K$
is equivalent to  mapping them to some feature space and applying the
dot-product. Therefore, each kernel defines an \emph{implicit} feature map.
This property is powerful: we can manipulate SQL queries as if they lived in a
vector space, but without actually materializing the space. In essence, kernel
methods offer a middle way between neighborhood-based approaches and feature
mapping.

\subsection{Kernels for the Query Log}
\label{sec:syntax}
In the past, authors have successfully built kernel functions for complex
objects, such as texts, DNA strings, \mbox{images} or even
videos~\cite{gartner2003survey}. Our task is now to design a kernel function
for SQL queries.

\textbf{Dissimilarity-Based Kernels.} Not all dissimilarity measures are kernel
functions. To qualify, a measure must obey \emph{Mercer's
conditions}~\cite{bishop2006pattern}. Those imply that the eigenvalues of the
dissimilarity matrix are positive. We know no function that guarantees these
conditions.  However, authors have presented methods to turn arbitrary
dissimilarity measures into kernels, such as \emph{spectral shifting} or
\emph{spectral clipping}~\cite{chen2009learning,wu2005analysis}.  These methods
compute the spectrum of the dissimilarity matrix, and correct the eigenvalues
to meet Mercer's conditions. In effect, they let us reuse the
dissimilarity measures from the literature, similarly to DBFMs. But they  are
costly, i.e., cubic with the number of items.  Also, it is not clear how to
maintain their results as new queries come in.

\textbf{Custom Kernels.} An alternative approach is to engineer new kernels
from scratch. Authors have developed such functions for graphs, sets, and even logic
programs~\cite{gartner2003survey}.  We could extend those to SQL queries. To
tackle different use cases, we could generate several kernels. For example, we
envision a function to describe the syntax of the queries, and another to
describe their set properties. We could easily aggregate them, because the
weighted sum of two kernels is itself a kernel.  But we could also attempt to
design a lossless solution. Indeed, kernels can encode infinite dimension
spaces. The Gaussian dissimilarity is a popular illustration of this
property~\cite{bishop2006pattern}.  Therefore, we do not exclude the existence
of a ``perfect'' kernel function for SQL queries.

\textbf{Discussion.} Compared to feature maps, kernel methods have many
advantages. They are possibly more space efficient, because they do not
materialize the vectors. The underlying encoding $\Phi$ is robust: it does not
involve arbitrary restrictions, and it is independent from the other queries in
the log. Lastly, kernels bypass the costly compression operations of feature
maps: the whole space is embedded in the dissimilarity function.

Nevertheless, our quest for a transformation $\Phi$ does not stop here. Even if
we had access to a perfect kernel, it is likely that its implicit feature space
would remain theoretical: we would know that the inverse feature map
$\Phi^{-1}$ exists, but we could not access it.  Also, not all statistical
methods were kernelized, hence kernel approaches are less general than explicit
methods. Finally, their accuracy for SQL log mining remains to be
studied.  In particular, we must evaluate their sensitivity to the curse of
dimensionality.

\section{Graphical Models}
\label{sec:generative}

\begin{figure}[!t]
  \centering
    \includegraphics[width=\columnwidth]{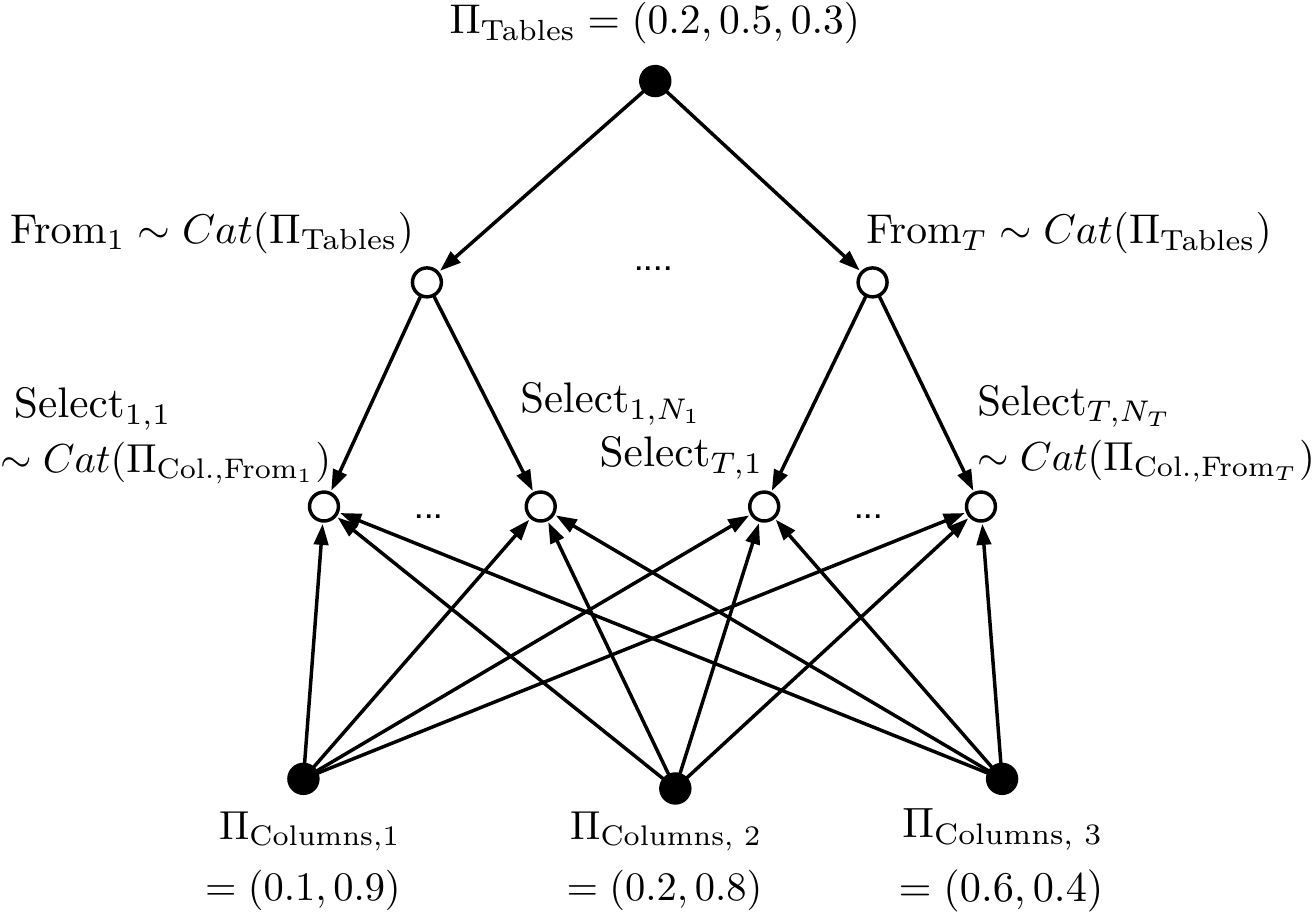}
    \caption{Simple Bayesian model to describe the distribution of
    \texttt{SELECT-FROM} queries, in a database made of three tables with two
columns each. The full circles represent constants, the empty circles represent
random variables.}
  \label{pic:bayesselectfrom}
\end{figure}

So far, we have only considered methods related to vector spaces. But there
exists an alternative conceptual framework for which many statistical
methods were developed: \emph{probabilistic graphical models}, also called
\emph{Bayesian networks}.

\textbf{Presentation.} The aim of graphical models is to decompose complex
probability distributions into elementary, low-dimension components. Let us
introduce an example. We wish to describe the distribution of all the
\texttt{SELECT-FROM} queries from the log of a DBMS. In other words, we want to
estimate the function $p_{SF}: \mathbb{Q}_{\texttt{SELECT-FROM}} \to [0,1]$,
which maps each query to its probability. Finding a closed mathematical form
for this function is difficult: it involves complex operations, many
parameters, and the number of these parameters is variable. Bayesian networks
give us a mean to express $p_{SF}$ in a graphical way.
Figure~\ref{pic:bayesselectfrom} displays an example of model. This graph can
be understood as an algorithm to generate new queries. We read it as follows:
\begin{itemize}
    \item Set the constant vectors  $\Pi_\text{Tables}$, $\Pi_{\text{Columns}, 1}$,
        $\Pi_{\text{Columns}, 2}$, and $\Pi_{\text{Columns}, 3}$. The vector
        $\Pi_\text{Tables}$ describes the probability of occurrence of all the
        tables. The vectors $\Pi_{\text{Columns}, t}$ describes the probability
        of occurrence of the columns in each table $t$.
    \item Chose $T$ random tables $\{\text{From}_1, \ldots, \text{From}_T\}$,
        picking them randomly with probabilities $\Pi_\text{Tables}$
    \item For each table $t \in \{\text{From}_1, \ldots, \text{From}_T\}$,
        chose $N_{t}$ random columns $\{\text{Select}_{t,1}, \ldots,
        \text{Select}_{t,N_t}\}$, picking randomly with probabilities
        $\Pi_{\text{Columns}, t}$.
\end{itemize}
Thus, the network describes a method to sample from the distribution
$p_{SF}$.  In fact, it also gives us a tractable way to compute the
probability $p_{SF}(Q)$ for any given query $Q$. Here again, we refer readers
to Bishop~\cite{bishop2006pattern} for more details.

\begin{figure}[!t]
  \centering
    \includegraphics[width=\columnwidth]{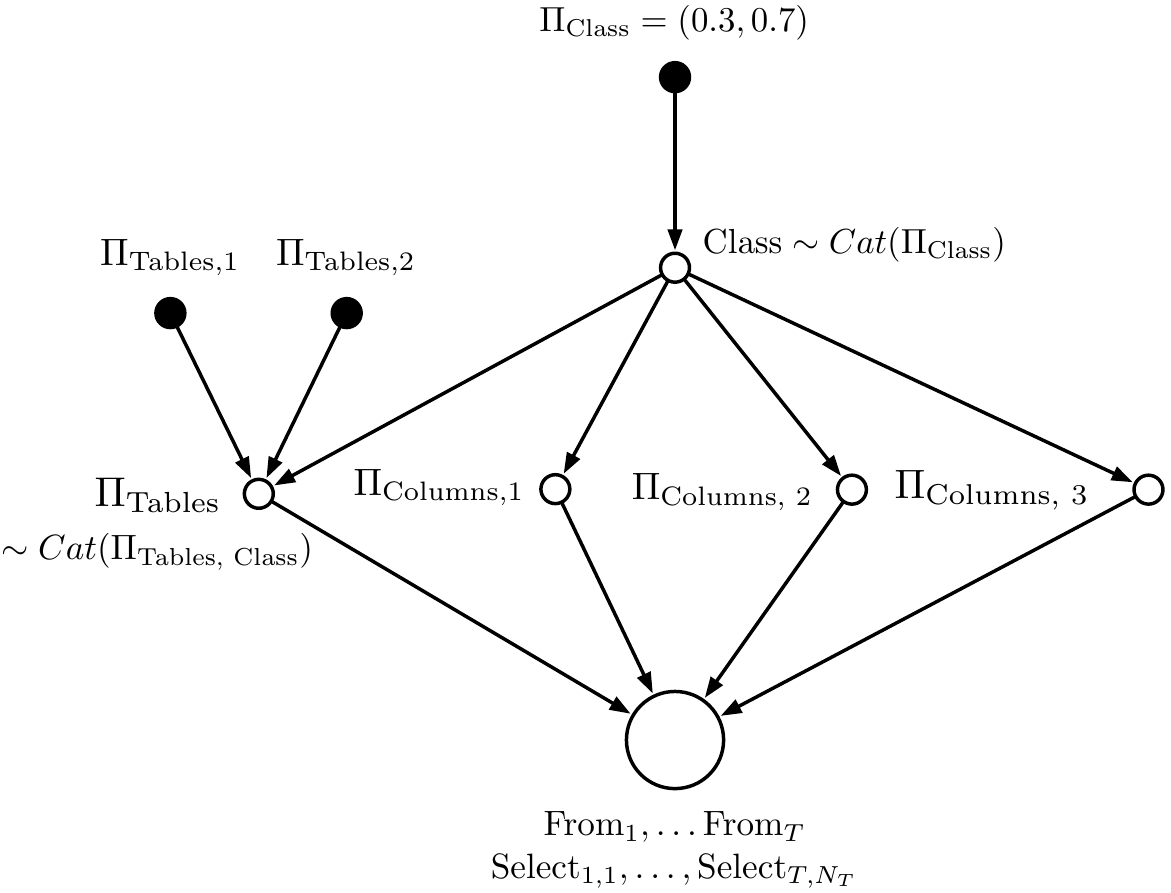}
    \caption{Extension of the \texttt{SELECT-FROM} model to support clusters.
        The latent variable $Class$ influences the distribution of the
        parameters $\Pi_\text{Tables}$, $\Pi_{\text{Columns}, 1}$,
        $\Pi_{\text{Columns}, 2}$, and $\Pi_{\text{Columns}, 3}$, which
        themselves influence the query, as illustrated in
        Figure~\ref{pic:bayesselectfrom}. The distribution of the variables
    $\Pi_{\text{Columns}, t}$ have form as that of $\Pi_\text{Tables}$, for $t
    \in\{1,2,3\}$.}
  \label{pic:bayesclustering}
\end{figure}
\textbf{Extensions.} With graphical models, we can compute complex probability
functions and generate samples. Accordingly, if we had a complete model for SQL
queries, we could detect ``typical'' or ``outlying'' queries, and we
could generate realistic SQL statements. But we could also extend the model to
cover more complex tasks. In the machine learning literature, authors have
described dozens of statistical methods with Bayesian networks, including all
those that interest us~\cite{bishop2006pattern}. We could exploit them, by
``plugging in'' our own SQL network. As an illustration, we present an
elementary clustering model in Figure~\ref{pic:bayesclustering}. To build this
model, we plugged our \texttt{SELECT-FROM} model into a mixture of
distributions. In Section~\ref{sec:bridge}, we will introduce more
general methods, to support all types of machine learning algorithms.

\textbf{Discussion.} Aside from dummy coding, Bayesian modeling is the only
method which provides both the mapping $\Phi$ and its inverse $\Phi^{-1}$.  To
obtain the image $\Phi(Q)$ of a given query $Q$, we instantiate the variables
in the network. To obtain its inverse $\Phi^{-1}(Q)$, we execute the generative
process. Additionally, graphical models are more flexible than vectors. For
instance, they support variable numbers of parameters and recursivity. Besides,
they are interpretable, and they have convenient statistical
properties: among \mbox{others}, Bayesian methods natively incorporate regularization and
adaptivity  (cf.  empirical Bayes~\cite{bishop2006pattern}).

Yet, producing a complete Bayesian network for SQL que\-ries remains a
challenge.  Also, adapting its parameters to the log may involve costly
computation methods, such as Monte-Carlo simulations. Finally, as with feature
maps and kernel functions, the empirical performance of this method remains to
be studied.  At this point, we do not know how accurate it is for log
mining.

\section{Bridging Graphical Models and Vector Spaces}
\label{sec:bridge}
\begin{figure}[!t]
  \centering
    \includegraphics[width=\columnwidth]{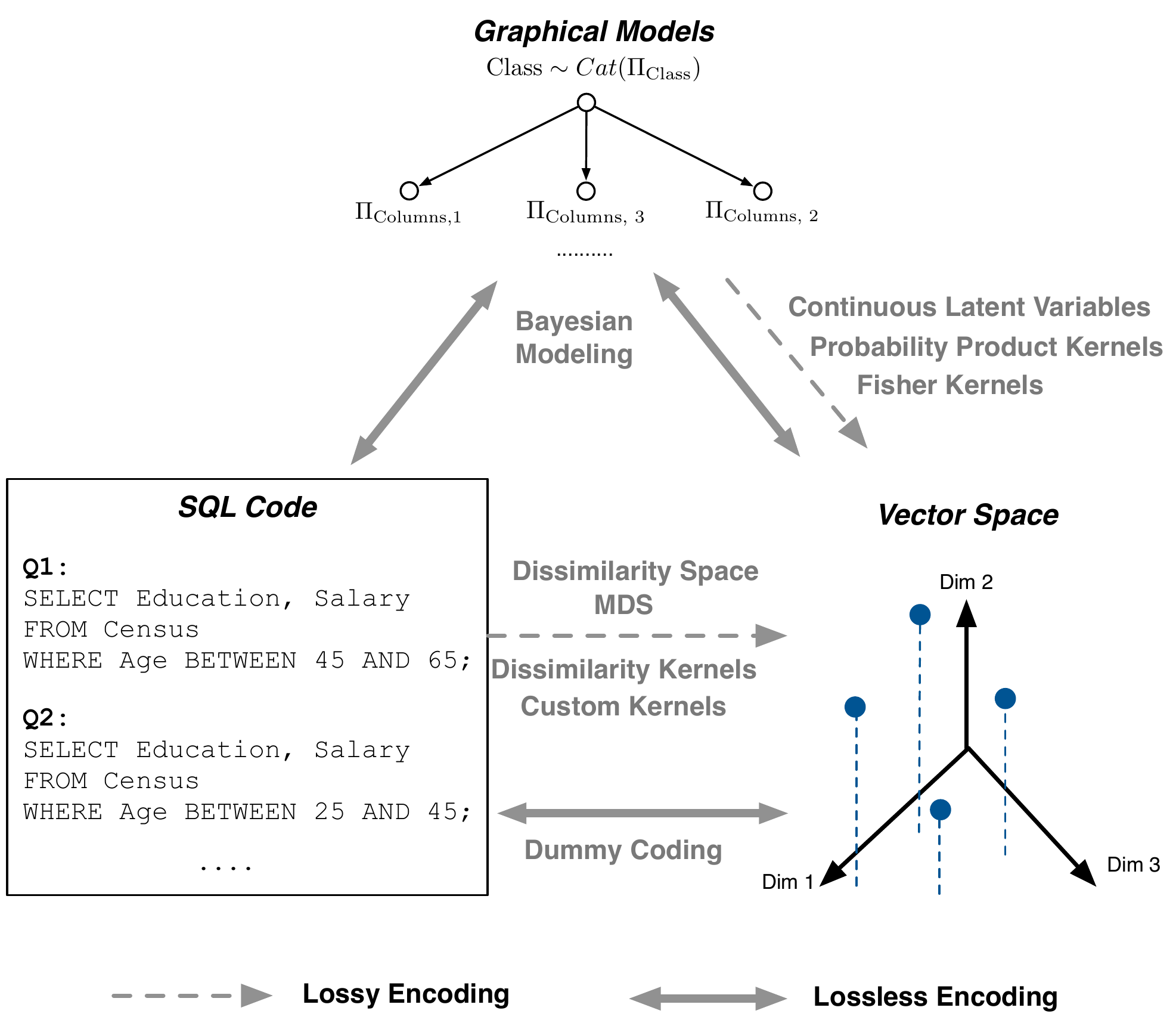}
    \caption{Summary of the methods discussed in this paper.}
  \label{pic:bayesselectfrom}
\end{figure}
To close our presentation, we highlight a powerful feature of probabilistic
graphical models: they can yield vector spaces, both implicitly and
explicitly.

For a start we can embed graphical models into kernel functions.  We know at
least two methods to do so, \emph{probabi\-li\-ty product kernels} and \emph{Fisher
kernels}~\cite{bishop2006pattern}. Thanks to these solutions, we can benefit
from both the generative features of graphical models and the libraries of
kernel methods.

Furthermore, we conjecture that we can generate vectors directly from graphical
models.  In Figure~\ref{pic:bayesclustering}, we show an example of latent
variable model, where the discrete variable $Cluster$ influences the
distribution of the query's components. We could generalize this model to
\emph{continuous latent variables}. In this case, a fixed-size random vector
would condition the distribution of the parameters $\Pi_\text{Tables}$ and
$\Pi_{\text{Columns}, t}$. The exact parametric form of the dependency has yet
to be determined.

Finally, observe that we can operate in the opposite direction, and convert
query-vectors $\phi_i$ into instances of a Bayesian network.  Several methods
exist to learn such models automatically from matrices. Nevertheless, their
practical interest is limited: we have no guarantee that the generated
graphical models will be complete, or interpretable. And they have no way to
recover the information destroyed by the feature maps.

We summarize all the methods in this paper and their relationships in
Figure~\ref{pic:bayesselectfrom}. Bayesian models seem to offer the ``best of
all worlds'':  they are lossless, reversible, and they can yield vector spaces.
For this reason, we chose to place them on top of our agenda. But we should not
underestimate their competitors. Even dummy coding may come in handy, in
conjunction with advanced compression algorithms such as autoencoders.  Now, our
task is to implement these ideas and conduct extensive benchmarks.  Eventually,
only practice and experiments will reveal which of these solutions truly fulfills
our vision.

\section{Related Work}
\label{sec:RelatedWork}

Several authors have developed methods to infer know\-ledge from the query log,
either to improve the performance of the database or to help users write
queries.

\textbf{Application-Specific Methods.} On the performance side, Ghosh et
al.~\cite{ghosh2002plan} associate each query from the log with a vector of
predefined scores (e.g., number of tables mentioned, number of joins, presence
of index) to recommend query plans.  Aouiche and Darmont~\cite{aouiche2009data}
mine the co\-lumn names mentioned in the log to chose materialized views and
indices.  The optimizer LEO~\cite{stillger2001leo} monitors the execution of
queries to predict cardinalities.  On the user side, Agrawal et
al.~\cite{agrawal2006context} have presented a method to recommend individual
tuples.  Yang et al.~\cite{yang2009recommending} mine the log for join
predicates.  SnipSuggest~\cite{khoussainova2010snipsuggest} suggests
context-sensitive snippets. Zhang has developed an interface to explore the
Sloan Digital Sky Survey database~\cite{zhang2011data}. Giacometti et
al.~\cite{giacometti2009query} present a method to detect unexpected patterns.
Finally, Yao et al.~\cite{yao2005finding} exploit cluster analysis to detect
so-called query sessions.

Each of these papers use a different, task-specific enco\-ding. Our ambition is
to develop one framework to encompass all those cases.

\textbf{Neighborhood-Based Methods.} We discuss these me\-thods in detail in our
introduction. We  generalize them with DBFMs, in Section~\ref{sec:explicit}.

\textbf{Hierarchical Modelling of Queries.} In Section~\ref{sec:generative}, we
present generative approaches. In fact, the early system
PROMISE~\cite{sapia2000promise},  based on Markov Models, is remarkably close
to our vision. However, it targets very specific OLAP workloads. SnipSuggest
also represents the queries with a tree~\cite{khoussainova2010snipsuggest}, but
the leaves represent fragments of plain text. Finally, the Oracle Workload
Intelligence also uses a Bayesian model~\cite{tran2015oracle}, but it operates
at the user session level: each node represents a complete query.

\textbf{Log Analysis in Information Retrieval.} Authors have developed many
methods to mine search engine query logs~\cite{silvestri2010mining}. In
principle, we could use those, exploiting natural language models such as
$n$-grams or tf-idf. But these me\-thods incur a major loss of information. First,
they neglect the grammar of SQL.  This is wasteful, because the language is
simple, highly structured, and well-known. Second, they neglect the
set relationships between the queries, such as inc\-lusion, overlap or
order. Those are crucial for many of the applications we target.

\section{Conclusion}
\label{sec:conclusion}

Too many methods to mine SQL query logs are isolated. They are isolated from
each other: each paper uses its own conventions and its own algorithms. They
are also isolated from the rest of machine learning research: they only exploit
a narrow subset of its literature. In this paper, we presented three research
directions to unify and broaden the scope of DBMS log mining. We purposely
stepped out of specific applications, and presented  frameworks to apply
general statistical inference on SQL queries.

We now envision two lines of research. First, we will implement all the methods
discussed in this paper, compare them, and understand which one performs best
and why. Once we have solid tools to encode SQL queries, we will experiment
with new machine learning algorithms. Given the recent advances in this field,
with e.g. deep learning, we are convinced that this agenda holds a bright
future.

\section{Acknowledgments}
This work was supported by the Dutch national program \\COMMIT.
\balance
\bibliographystyle{abbrv}
\balance
\bibliography{SQLkernel}

\begin{thebibliography}{10}



\bibitem{agrawal2006context}
R.~Agrawal, R.~Rantzau, and E.~Terzi.
\newblock Context-sensitive ranking.
\newblock In {\em Proc. SIGMOD}, pages 383--394, 2006.

\bibitem{akbarnejad2010sql}
J.~Akbarnejad, M.~Eirinaki, S.~Koshy, D.~On, and N.~Polyzotis.
\newblock Sql querie recommendations: a query fragment-based approach.
\newblock {\em Proc. VLDB}, 2010.

\bibitem{aligon2014similarity}
J.~Aligon, M.~Golfarelli, P.~Marcel, S.~Rizzi, and E.~Turricchia.
\newblock Similarity measures for olap sessions.
\newblock {\em Knowledge and Information Systems}, pages 463--489, 2014.

\bibitem{aouiche2009data}
K.~Aouiche and J.~Darmont.
\newblock Data mining-based materialized view and index selection in data
  warehouses.
\newblock {\em Journal of Intelligent Information Systems}, pages 65--93, 2009.

\bibitem{bishop2006pattern}
C.~M. Bishop.
\newblock {\em Pattern recognition and machine learning}.
\newblock Springer, 2006.

\bibitem{borg2005modern}
I.~Borg and P.~J. Groenen.
\newblock {\em Modern Multidimensional Scaling: Theory and Applications}.
\newblock Springer, 2005.

\bibitem{chatzopoulou2009query}
G.~Chatzopoulou, M.~Eirinaki, and N.~Polyzotis.
\newblock Query recommendations for interactive database exploration.
\newblock In {\em SSDBM}, 2009.

\bibitem{chen2009learning}
Y.~Chen, M.~R. Gupta, and B.~Recht.
\newblock Learning kernels from indefinite similarities.
\newblock In {\em Proc. ICML}, pages 145--152, 2009.

\bibitem{desrosiers2011comprehensive}
C.~Desrosiers and G.~Karypis.
\newblock A comprehensive survey of neighborhood-based recommendation methods.
\newblock {\em Recommender Systems Handbook}, pages 107--144, 2011.

\bibitem{duin2012dissimilarity}
R.~P. Duin and E.~Pekalska.
\newblock The dissimilarity space: Bridging structural and statistical pattern
  recognition.
\newblock {\em Pattern Recognition Letters}, pages 826--832, 2012.

\bibitem{gartner2003survey}
T.~G{\"a}rtner.
\newblock A survey of kernels for structured data.
\newblock {\em SIGKDD Explorations}, pages 49--58, 2003.

\bibitem{ghosh2002plan}
A.~Ghosh, J.~Parikh, V.~S. Sengar, and J.~R. Haritsa.
\newblock Plan selection based on query clustering.
\newblock In {\em Proc. VLDB}, pages 179--190, 2002.

\bibitem{giacometti2009query}
A.~Giacometti, P.~Marcel, E.~Negre, and A.~Soulet.
\newblock Query recommendations for olap discovery driven analysis.
\newblock In {\em Proc. DOLAP}, pages 81--88, 2009.

\bibitem{halim2012stochastic}
F.~Halim, S.~Idreos, P.~Karras, and R.~H. Yap.
\newblock Stochastic database cracking: Towards robust adaptive indexing in
  main-memory column-stores.
\newblock {\em Proc. VLDB}, pages 502--513, 2012.

\bibitem{khoussainova2010snipsuggest}
N.~Khoussainova, Y.~Kwon, M.~Balazinska, and D.~Suciu.
\newblock Snipsuggest: context-aware autocompletion for sql.
\newblock {\em Proc. VLDB}, 2010.

\bibitem{koren2008factorization}
Y.~Koren.
\newblock Factorization meets the neighborhood: a multifaceted collaborative
  filtering model.
\newblock In {\em Proc. SIGKDD}, pages 426--434, 2008.

\bibitem{Nguyen2015Ident}
H.~V. Nguyen, K.~B{\"{o}}hm, F.~Becker, B.~Goldman, G.~Hinkel, and
  E.~M{\"{u}}ller.
\newblock Identifying user interests within the data space - a case study with
  skyserver.
\newblock In {\em Proc. EDBT}, pages 641--652, 2015.

\bibitem{sapia2000promise}
C.~Sapia.
\newblock Promise: Predicting query behavior to enable predictive caching
  strategies for olap systems.
\newblock In {\em Proc. DaWaK}, pages 224--233, 2000.

\bibitem{sarawagi1998discovery}
S.~Sarawagi, R.~Agrawal, and N.~Megiddo.
\newblock Discovery-driven exploration of olap data cubes.
\newblock {\em Proc. EDBT}, 1998.

\bibitem{silvestri2010mining}
F.~Silvestri.
\newblock Mining query logs: Turning search usage data into knowledge.
\newblock {\em Foundations and Trends in Information Retrieval}, pages 1--174,
  2010.

\bibitem{stillger2001leo}
M.~Stillger, G.~M. Lohman, V.~Markl, and M.~Kandil.
\newblock Leo-db2's learning optimizer.
\newblock In {\em Proc. VLDB}, pages 19--28, 2001.

\bibitem{tran2015oracle}
Q.~T. Tran, K.~Morfonios, and N.~Polyzotis.
\newblock Oracle workload intelligence.
\newblock In {\em Proc. SIGMOD}, pages 1669--1681, 2015.

\bibitem{wu2005analysis}
G.~Wu, E.~Y. Chang, and Z.~Zhang.
\newblock An analysis of transformation on non-positive semidefinite similarity
  matrix for kernel machines.
\newblock In {\em Proc. ICML}, 2005.

\bibitem{yang2009recommending}
X.~Yang, C.~M. Procopiuc, and D.~Srivastava.
\newblock Recommending join queries via query log analysis.
\newblock In {\em Proc. ICDE}, pages 964--975. IEEE, 2009.

\bibitem{yao2005finding}
Q.~Yao, A.~An, and X.~Huang.
\newblock Finding and analyzing database user sessions.
\newblock In {\em Proc. DASFAA}, pages 851--862, 2005.

\bibitem{zhang2011data}
J.~Zhang.
\newblock {\em Data Use and Access Behavior in eScience---Exploring data
  practices in the new data-intensive science paradigm}.
\newblock PhD thesis, Drexel University, 2011.

\end{thebibliography}

\end{document}